\documentclass{blois}

\def\Journal#1#2#3#4{{#1} {\bf #2}, #3 (#4)}

\def\NIMA{{\em Nucl. Instrum. Methods} A}

\def\PLB{{\em Phys. Lett.}  B}
\def\PRL{\em Phys. Rev. Lett.}
\def\PRD{{\em Phys. Rev.} D}

\def\JHEP{{\em J. High Energy Phys.}}
\def\PTEP{{\em Prog. Theor. Exp. Phys.}}
\def\EPJC{{\em Eur. Phys. J.} C}
\def\NIMA{{\em Nucl. Instrum. Meth.} A}

\def\be{\begin{equation}}
\def\ee{\end{equation}}
\def\bea{\begin{eqnarray}}
\def\eea{\end{eqnarray}}

\usepackage{amsmath}
\usepackage{setspace}

\begin{document}
\vspace*{4cm}
\title{Recent Belle II Results on Hadronic $\boldsymbol{B}$ Decays}

\author{Shu-Ping Lin, on behalf of the Belle II Collaboration}

\address{Department of Physics, National Taiwan University, \\
No. 1, Sec. 4, Roosevelt Rd., Taipei 106216, Taiwan}

\maketitle
\abstracts{
The investigation of $B$ meson decays to charmed and charmless hadronic final states is a keystone of the Belle II program. 
Analyses of such decays provide reliable and experimentally precise constraints on the weak interactions of quarks. They are sensitive to effects from non-SM physics, and further our knowledge about uncharted $b\rightarrow c$ hadronic transitions.
We present new results from combined analyses of Belle and Belle II data to determine the quark-mixing parameter $\phi_3$ (or $\gamma$), and from the Belle II analyses of two-body decays that are related to the determination of $\phi_2$ (or $\alpha$).
We also present recent Belle II results on branching ratios and direct {\it CP}-violating asymmetries of several $B$ decays, which result in a competitive standard-model test based on the $K\pi$ isospin sum rule and first observations of three new $B \rightarrow D^{(*)} K K_S^0$ decays. 
}

\section{Introduction}
Hadronic $B$ decays provide precise constraints on the parameters of the Cabibbo-Kobayashi-Maskawa (CKM) quark-mixing matrix and are sensitive probes of physics beyond the standard model (SM).
Measurement of new decay channels expands our knowledge of the flavour sector. 
Belle II~\cite{BelleII}, located at the SuperKEKB~\cite{SuperKEKB} asymmetric collider at KEK, is a hermetic solenoidal magnetic spectrometer surrounded by particle-identification detectors, an electromagnetic calorimeter, and muon detectors. Optimised for the reconstruction of bottom-antibottom pairs produced at the threshold in $\Upsilon(4S)$ decays, Belle II has competitive, unique, or world-leading reach in many key quantities associated with hadronic $B$ decays.
The CKM angle $\phi_3/\gamma = \arg\left( - \frac{V_{ud}V^*_{ub}}{V_{cd}V_{cb}^*} \right)$ is a fundamental constraint for charge-parity ({\it CP}) violation in the SM and can be reliably determined in tree-level processes, with negligible loop-amplitude contributions.
The angle $\phi_2/\alpha = \arg\left( - \frac{V_{td}V^*_{tb}}{V_{ud}V_{ub}^*} \right)$ is currently limiting the precision of non-SM tests based on global fits of the quark-mixing matrix unitarity and is best determined through measurement of $B\rightarrow\rho\rho$ and $B\rightarrow\pi\pi$ decays.
For $B\rightarrow K\pi$ decays, an isospin sum rule~\cite{sum_rule} combines the branching fractions and {\it CP} asymmetries of the decays, providing a null test of the SM.
The capability of investigating all the related final states jointly, under consistent experimental settings, is unique to Belle II.

The challenge in analysing these channels lies in the large amount of $e^+e^-\rightarrow q\bar{q}$ continuum background. Binary-decision-tree classifiers $C$ are used to discriminate between signal and continuum events using event topology, kinematic, and decay-length information.~\cite{BDT} The determination of the signal yields is mainly based on observables that exploit the specific properties of at-threshold production: the energy difference between the $B$ candidate and the beam energy, $\Delta E = E_B^* - E_{\text{beam}}^*$, and the beam-constrained mass $M_{\text{bc}} = \sqrt{E_\text{beam}^*/c^2-(\boldsymbol{p_B^*}/c)^2}$, where $E_B^*$ and $\boldsymbol{p_B^*}$ are the energy and momentum of the $B$ candidate, respectively, and $E_{\text{beam}}^*$ is the beam energy, all in the center-of-mass frame.

\section{$\boldsymbol{B \rightarrow D^{(*)} K K_S^0}$ Decays}
\begin{figure}[htb]
    \centering
    \includegraphics[width=\linewidth]{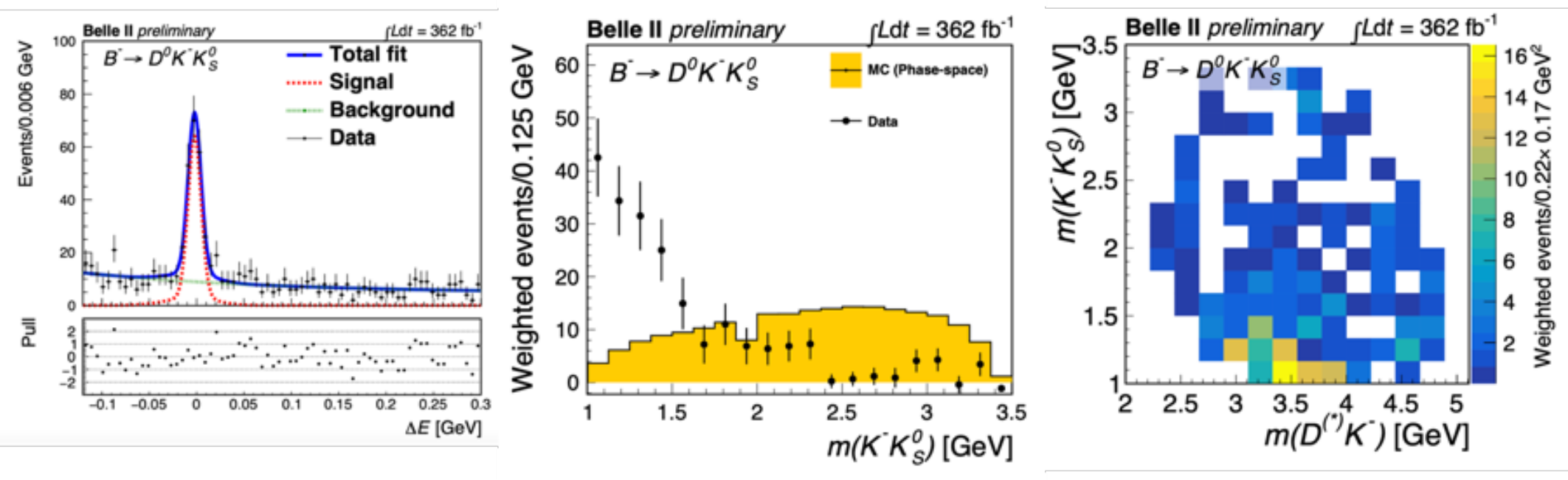}
    \caption{$\Delta E$ (left), $m(KK_S^0)$ (middle), and Dalitz (right) distributions of of $B^- \rightarrow D^0K^-K_S^0$ decay.}
    \label{fig:DKK}
\end{figure}

The composition of a large fraction of the $B$ hadronic width is unknown, which limits significantly our capability to model $B$ decays in simulation, impacting the precision of many measurements. Belle II is pursuing a systematic program of exploration of hadronic $B$ decays. 
We report a preliminary measurement of the branching fraction of $B^- \rightarrow D^0K^-K_S^0$ decay~\cite{DKK_Belle2}, with a precision that is three times better than the previous best results~\cite{DKK_Belle}, 
and first observation of three new decay channels ($\bar{B}^0 \rightarrow D^+K^-K_S^0$, $B^- \rightarrow D^{*0}K^-K_S^0$, $\bar{B}^0 \rightarrow D^{*+}K^-K_S^0$)~\cite{DKK_Belle2}.
The branching fractions
\begin{align*}
\mathcal{B}(B^- \rightarrow D^0 K^- K_S^0) &= (1.89 \pm 0.16 \pm 0.10) \times 10^{-4}, \\
\mathcal{B}(\bar{B}^0 \rightarrow D^+ K^- K_S^0) &= (0.85 \pm 0.11 \pm 0.05) \times 10^{-4}, \\
\mathcal{B}(B^- \rightarrow D^{*0} K^- K_S^0) &= (1.57 \pm 0.27 \pm 0.12) \times 10^{-4}, \\
\mathcal{B}(\bar{B}^0 \rightarrow D^{*+} K^- K_S^0) &= (0.96 \pm 0.18 \pm 0.06) \times 10^{-4}, 
\end{align*}
are extracted from a $362\text{ fb}^{-1}$ Belle II sample using likelihood fits to the unbinned distributions of the energy difference $\Delta E$,
where the first uncertainties are statistical and the second are systematic.
The invariant mass $m(KK_S^0)$ of the two kaons is investigated. 
For all four channels, the $m(KK_S^0)$ distribution exhibits a peaking structure in the low-mass region, which departs from the expected three-body phase space distribution. 
Structures are also observed in the Dalitz distributions (Figure~\ref{fig:DKK}).

\section{Determination of CKM Angle $\boldsymbol{\phi_3/\gamma}$}
The angle $\phi_3$ is studied through the interference of $b\rightarrow c\bar{u} s$ and $b\rightarrow u\bar{c} s$ transition amplitudes in tree-level $B$ hadron decays. 
The current world average $\phi_3=\left( 65.9^{+3.3}_{-3.5}\right)^\circ$ is dominated by LHCb measurements~\cite{PDG_2022,phi3_LHCb}.
$\phi_3$ can be determined through different approaches featuring different $D$ final states from $B$ decays into charmed $B$ final states.
The most precise Belle II result is obtained with self-conjugate $D$ final states ($D \rightarrow K_S^0 h^+ h^-$, $h=K,\pi$)~\cite{phi3_belle2_jhep}, where several intermediate resonances are involved in $D$ decays, resulting in variation of the {\it CP} asymmetry over the phase space.
Two other approaches have also been pursued: one with Cabibbo-suppressed $D$ decays, and the other with the $D$ meson decaying to two-body {\it CP} eigenstates.
In both analyses it is required that $M_\text{bc} > 5.27 \text{ GeV}/c^2$, $|\Delta E| < 0.15 \text{ GeV}$ ($\Delta E>-0.13 \text{ GeV}$ for the latter), and a loose requirement on $C$ removes about 60\% of the continuum background. The signal yields are extracted with fits of $\Delta E$ and the continuum suppression classifier.

\subsection{Cabibbo-suppressed channels}
Grossman, Ligeti, and Soffer (GLS)~\cite{GLS_method} proposed a method to measure $\phi_3$ with singly Cabibbo-suppressed decays of $D$ mesons, $B^{\pm} \rightarrow D(\rightarrow K_S^0 K^{\pm} \pi^{\mp}) h^{\pm}$, where $D$ is the superposition of $D^0$ and $\bar{D}^0$ mesons. The $B^{\pm}$ meson can have the same sign (SS) or opposite sign (OS) with respect to the $K^{\pm}$ from the $D$ decay.
In this analysis~\cite{phi3_GLS_Belle2}, the information about the dynamics of the $D$ decays is integrated with external inputs from CLEO~\cite{D_cleo}.

The fitting is performed in both the full $D$ phase space and in the $K^*$ region, where the invariant mass $m(KK_S^0)$ is close to that of $K^*(892)^0$, thus enhancing the interference and the precision on $\phi_3$, due to the large strong-phase difference in $D\rightarrow K_S^0 K^{\pm} \pi^{\mp}$ decays.
Combining Belle ($711\text{ fb}^{-1}$) and Belle II ($362\text{ fb}^{-1}$) data, the result is consistent, though not yet competitive to the result from LHCb~\cite{phi3_GLS_LHCb}.
It can provide a constraint on $\phi_3$ when combined with results from other measurements.

\begin{table}[htb]
    \centering
    \onehalfspacing
    \caption{Results of $B^{\pm} \rightarrow D(\rightarrow K_S^0 K^{\pm} \pi^{\mp}) h^{\pm}$ decay combining Belle ($711\text{ fb}^{-1}$) and Belle II ($362\text{ fb}^{-1}$) data for the full $D$ phase space and the $K^*$ region. The first uncertainties are statistical and the second are systematic.}
    \label{tab:GLS_result}
    \vspace{0.2cm}
    \begin{tabular}{|l|c|c|}
    \hline
     & Full $D$ phase space & $K^*$ region \\
    \hline
    $\mathcal{A}^{DK}_{SS}$ & $-0.089 \pm 0.091 \pm 0.011$ & $0.055 \pm 0.119 \pm 0.020$ \\
    $\mathcal{A}^{DK}_{OS}$ & $0.109 \pm 0.133 \pm 0.013$ & $0.231 \pm 0.184 \pm 0.014$ \\
    $\mathcal{A}^{D\pi}_{SS}$ & $0.018 \pm 0.026 \pm 0.009$ & $0.046 \pm 0.029 \pm 0.016$\\
    $\mathcal{A}^{D\pi}_{OS}$ & $-0.028 \pm 0.031 \pm 0.009$ & $0.009 \pm 0.046 \pm 0.009$ \\
    $\mathcal{R}_{SS}^{DK/D\pi}$ & $0.122 \pm 0.012 \pm 0.004$ & $0.093 \pm 0.012 \pm 0.005$ \\
    $\mathcal{R}_{OS}^{DK/D\pi}$ & $0.093 \pm 0.013 \pm 0.003$ & $0.103 \pm 0.020 \pm 0.006$ \\
    $\mathcal{R}_{SS/OS}^{D\pi}$ & $1.428 \pm 0.057 \pm 0.002$ & $2.412 \pm 0.132 \pm 0.019$ \\
    \hline
    \end{tabular}
\end{table}

\subsection{{\it CP} eigenstates}
Gronau, London, and Wyler (GLW)~\cite{GL_method,GW_method} proposed a method where the $D$ meson decays to $K^+K^-$ ({\it CP}-even) or $K_S^0\pi^0$ ({\it CP}-odd) eigenstates. The $K_S^0\pi^0$ eigenstate has been accessible to $B$-factories only.
Belle ($711\text{ fb}^{-1}$) and Belle II ($189\text{ fb}^{-1}$) data are combined to give the final result, which is consistent but not competitive with results from BaBar~\cite{phi3_GLW_BaBar} and LHCb~\cite{phi3_GLW_LHCb}.
The branching fraction ratios ($\mathcal{R}_{{\it CP}\pm}$) and {\it CP} asymmetries ($\mathcal{A}_{{\it CP}\pm}$) are 
\begin{align*}
\mathcal{R}_{CP+} = 1.164 \pm 0.081 \pm 0.036, \\
\mathcal{R}_{CP-} = 1.151 \pm 0.074 \pm 0.019, \\
\mathcal{A}_{CP+} = 0.125 \pm 0.058 \pm 0.014, \\
\mathcal{A}_{CP-} = -0.167 \pm 0.057 \pm 0.060, 
\end{align*}
where the first uncertainties are statistical and the second are systematic.
The statistical and systematic precisions are significantly better than the previous Belle measurement~\cite{phi3_GLW_Belle}, and the result could constrain $\phi_3$ in combination with other measurements.
In particular, the first evidence for $\mathcal{A}_{{\it CP}+}$ and $\mathcal{A}_{{\it CP}-}$ having opposite signs is observed, showing clearly the effect of {\it CP} violation.

\section{Toward CKM Angle $\boldsymbol{\phi_2/\alpha}$}
The least precisely known CKM angle $\phi_2/\alpha$ is starting to limit the testing power of the CKM model.
Belle II has the unique capability of measuring all $B\rightarrow\rho\rho$ and $B\rightarrow\pi\pi$ decays, from which $\phi_2$ can be determined. The combined information from these decays exploits isospin symmetry, reducing the effect of hadronic uncertainties.
The $B$ candidates are required to have $M_\text{bc} > 5.27 \text{ GeV}/c^2$ and $|\Delta E| < 0.15 \text{ GeV}$ ($<0.30 \text{ GeV}$ for final states involving neutral pions), followed by continuum suppression that removes 90-99\% of continuum background.

\subsection{$B\rightarrow\rho\rho$ decays}
The measurement of $B\rightarrow\rho\rho$ decays require a complex angular analysis. The fit is based on $M_\text{bc}$, $\Delta E$, the dipion masses, and the helicity angle of the $\rho$ candidates.
The preliminary Belle II results of $B^0\rightarrow\rho^+\rho^-$ and $B^+\rightarrow\rho^+\rho^0$ decays using $189\text{ fb}^{-1}$ of data~\cite{rho+rho-,rho+rho0} are on par with the best performances from Belle~\cite{rho+rho-_Belle_2,rho+rho0_Belle} and BaBar~\cite{rho+rho-_BaBar,rho+rho0_BaBar}.
Results are listed in Table~\ref{tab:rhorho}.

\subsection{$B\rightarrow\pi\pi$ decays}
Measurements of $B^0\rightarrow\pi^+\pi^-$ and $B^+\rightarrow\pi^+\pi^0$ decays are based on $362\text{ fb}^{-1}$ of Belle II data. The fit is performed over $\Delta E$ and $C'$, and the dominant systematic uncertainty arises from the $\pi^0$ efficiency.
The first measurement of $B^0\rightarrow\pi^0\pi^0$ at Belle II is also reported~\cite{pi0pi0_Belle2}, using $189\text{ fb}^{-1}$ of data. This decay is both CKM- and colour-suppressed, and has only photons in the final state, making it experimentally challenging to measure. The result obtained from a fit to $M_\text{bc}$, $\Delta E$, and $C'$, however, achieves Belle's precision despite using a dataset that is only one third of the size. This is due to the dedicated $\pi^0$ selection and continuum suppression studies that yield a much higher $\pi^0$ efficiency.
Results are listed in Table~\ref{tab:rhorho}.

\begin{table}[htb]
    \centering
    \onehalfspacing
    \caption{$B\rightarrow\rho\rho$ and $B\rightarrow\pi\pi$ results. The first uncertainties are statistical and the second are systematic.}
    \label{tab:rhorho}
    \vspace{0.2cm}
    \begin{tabular}{|l|c|c|c|}
    \hline
    & $\mathcal{B}\ [10^{-6}]$ & $\mathcal{A}_{{\it CP}}$ & $f_L$ \\
    \hline
    $B^0\rightarrow\rho^+\rho^-$ & $26.7 \pm 2.8 \pm 2.8$ & -- & $0.956 \pm 0.035 \pm 0.033$ \\
    $B^+\rightarrow\rho^+\rho^0$ & $23.2 ^{+2.2}_{-2.1} \pm 2.7$ & $-0.069 \pm 0.068 \pm 0.060$ & $0.943 ^{+0.035} _{-0.033} \pm 0.027$ \\
    \hline
    $B^0\rightarrow\pi^+\pi^-$ & $5.83 \pm 0.22 \pm 0.17$ & -- & -- \\
    $B^+\rightarrow\pi^+\pi^0$ & $5.02 \pm 0.28 \pm 0.31$ & $-0.082 \pm 0.054\pm 0.008$ & -- \\
    $B^0\rightarrow\pi^0\pi^0$ & $1.38\pm 0.27 \pm 0.22$ & $0.14 \pm 0.46\pm 0.07$ & -- \\
    \hline
    \end{tabular}
\end{table}

\section{$\boldsymbol{K\pi}$ Isospin Sum Rule}
The isospin sum rule $I_{K\pi}$ is defined by
\begin{equation}
I_{K\pi} =  \mathcal{A}_{K^+\pi^-} + \mathcal{A}_{K^0\pi^+} \cdot \frac{\mathcal{B}_{K^0\pi^+}}{\mathcal{B}_{K^+\pi^-}}\frac{\tau_{B^0}}{\tau_{B^+}} 
  - 2\mathcal{A}_{K^+\pi^0} \cdot \frac{\mathcal{B}_{K^+\pi^0}}{\mathcal{B}_{K^+\pi^-}}\frac{\tau_{B^0}}{\tau_{B^+}} 
 -2\mathcal{A}_{K^0\pi^0}
\cdot\frac{\mathcal{B}_{K^0\pi^0}}{\mathcal{B}_{K^+\pi^-}},
\end{equation}
where $\mathcal{B}_{K\pi}$ and $\mathcal{A}_{K\pi}$ are the branching fractions and the {\it CP} asymmetries, and $\tau_{B^0}/\tau_{B^+} = 0.9273 \pm 0.0033$~\cite{PDG_2022} is the ratio of $B^0$ and $B^+$ lifetimes. The SM prediction of the sum rule is zero, with a precision of better than 1\%, in the limit of isospin symmetry and no electroweak penguins contributions.
Any large deviation from the SM prediction is an indication of non-SM physics.
The experimental precision of the sum rule is limited by  $\mathcal{A}_{K^0\pi^0}$~\cite{PDG_2022}.

\begin{figure}[htb]
    \centering
    \includegraphics[width=0.49\linewidth]{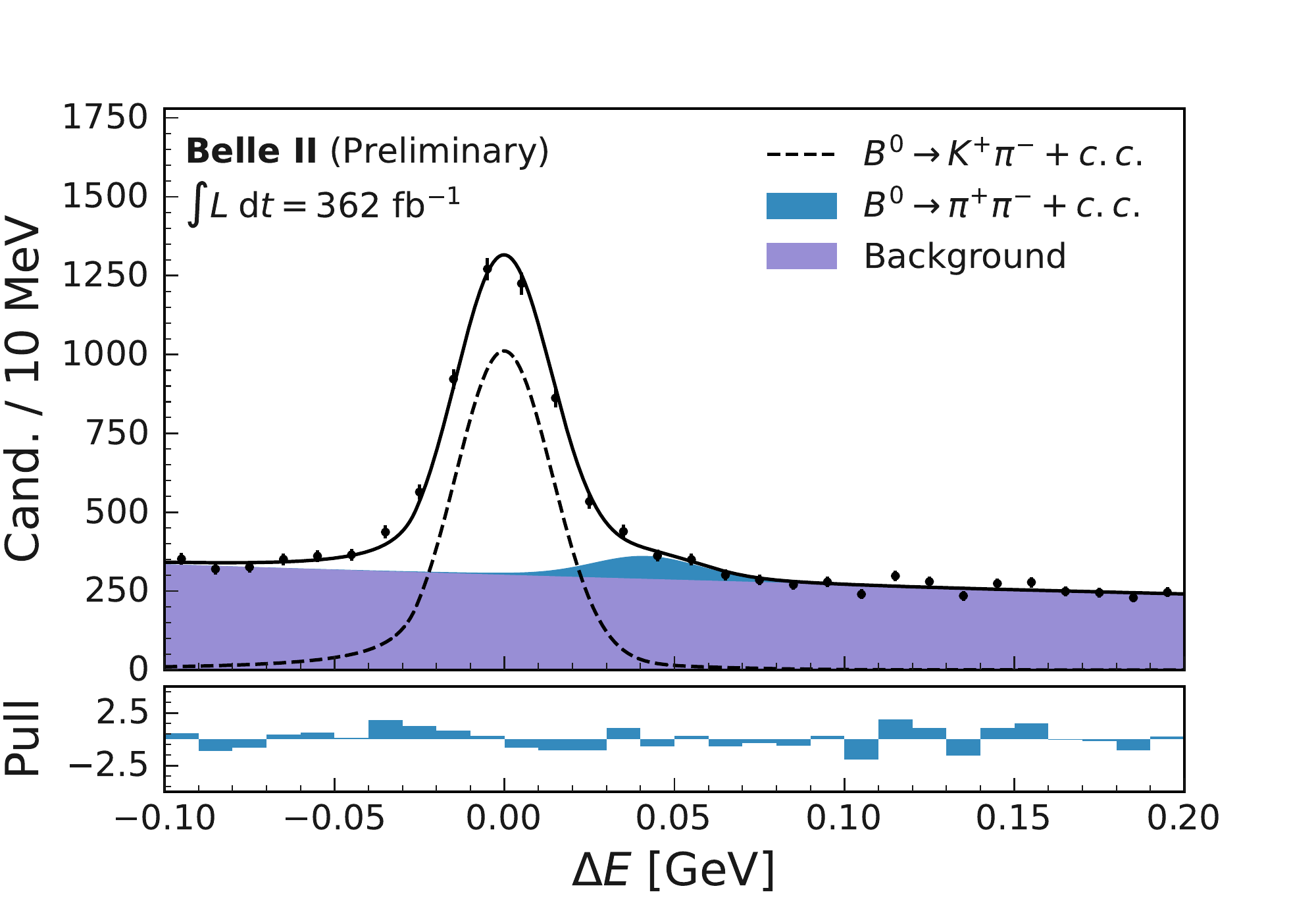}
    \includegraphics[width=0.49\linewidth]{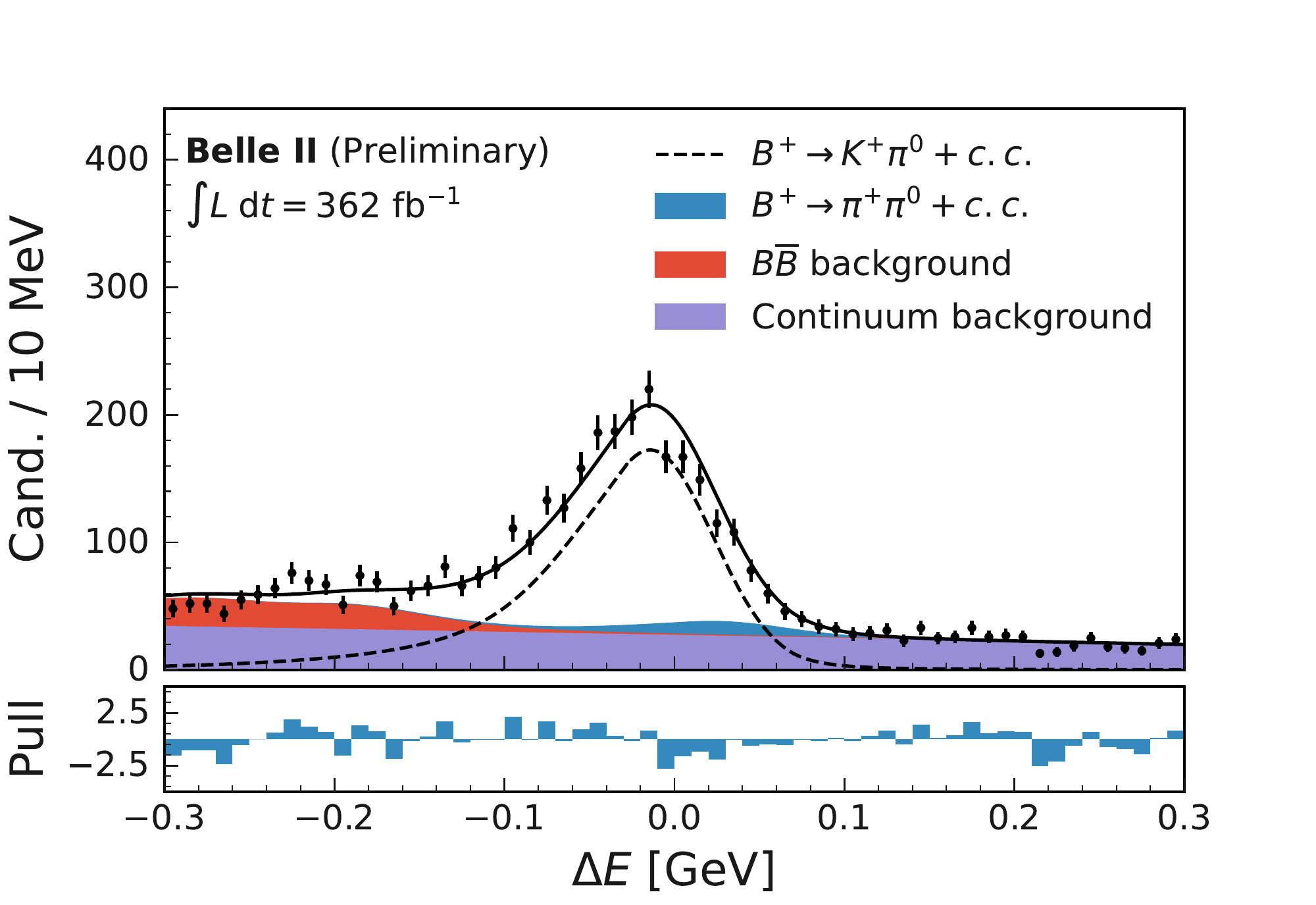}
    \includegraphics[width=0.49\linewidth]{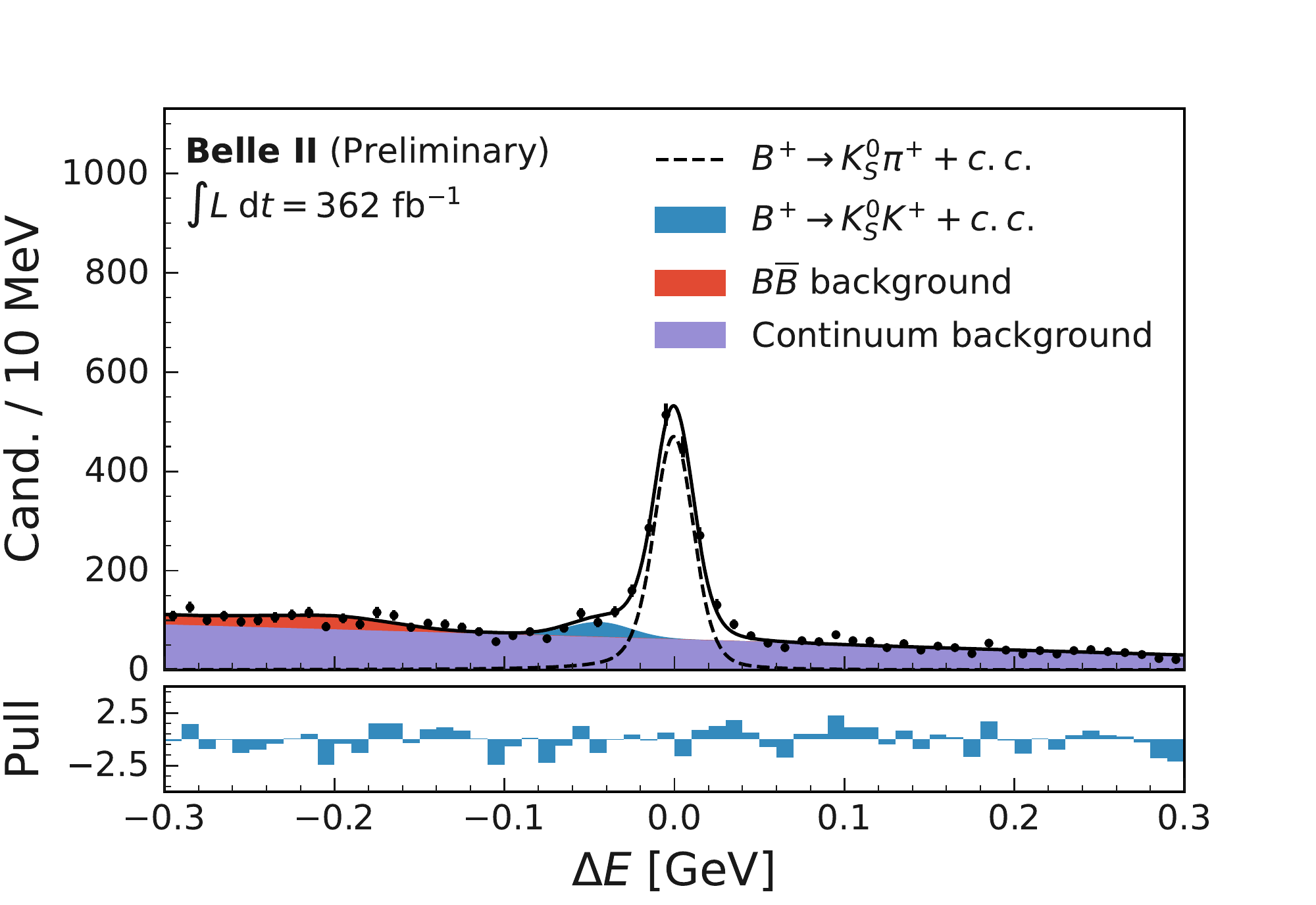}
    \includegraphics[width=0.49\linewidth]{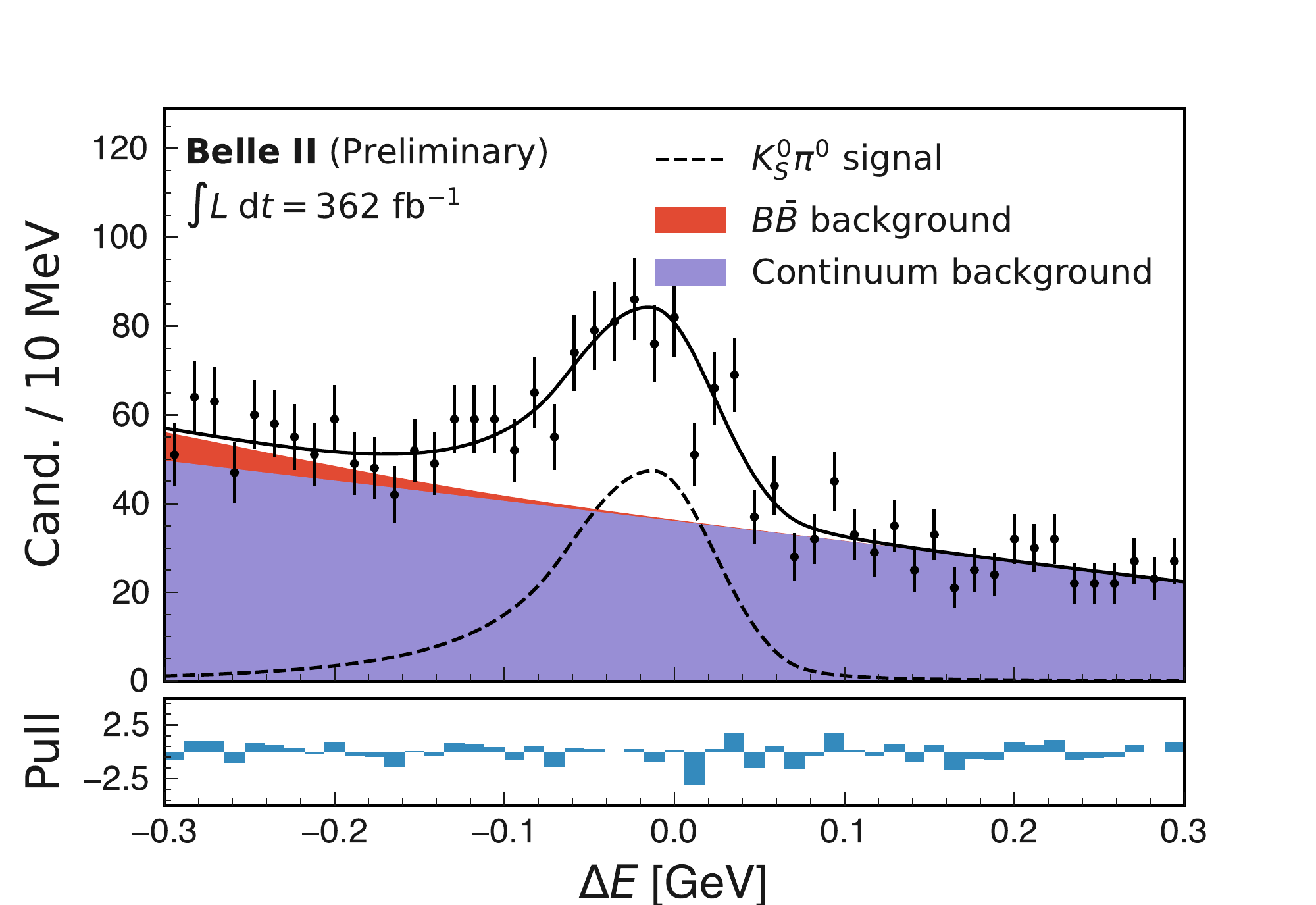}
    \caption{$\Delta E$ distributions of $B^0\rightarrow K^+\pi^-$ (upper left), $B^+\rightarrow K^+\pi^0$ (upper right), $B^+\rightarrow K_S^0\pi^+$ (lower left) and $B^0\rightarrow K_S^0\pi^0$ (lower right) decays.}
    \label{fig:Kpi}
\end{figure}

We studied all the final states associated with the sum rule: $B^0\rightarrow K^+\pi^-$, $B^+\rightarrow K_S^0\pi^+$, $B^+\rightarrow K^+\pi^0$, and $B^0\rightarrow K_S^0\pi^0$ using $362\text{ fb}^{-1}$ of Belle II data.
The analyses of the various decays follow a similar strategy, with common selections applied to the final states particles.
$B$ candidates are required to satisfy $5.272 < M_\text{bc} < 5.288 \text{ GeV}/c^2$, $|\Delta E| < 0.3 \text{ GeV}$, and a loose requirement of $C$ that suppresses 90-99\% of continuum background.
A fit is performed on the $\Delta E$-$C'$ distribution, where the flavour tagging algorithm~\cite{flv_tagger} is employed to determine the flavour of the $B$ candidate in $B^0\rightarrow K_S^0\pi^0$ decay due to absence of primary charged particles.
The $\Delta E$ distributions are shown in Figure~\ref{fig:Kpi}.
The measured branching fractions and {\it CP} asymmetries, as well as the sum rule calculated using these measurements, are listed in Table~\ref{tab:Kpi}.
They agree with the world averages~\cite{PDG_2022} and have competitive precisions.
In particular, the time-integrated and time-dependent results of $B^0\rightarrow K_S^0\pi^0$ are combined to achieve the world's best result for $\mathcal{A}_{K^0\pi^0}$, and consequentially for $I_{K\pi}$ a competitive precision that is limited by the statistical uncertainty.

\begin{table}[htb]
    \centering
    \onehalfspacing
    \caption{$B\rightarrow K\pi$ results using $362\text{ fb}^{-1}$ Belle II data. The first uncertainties are statistical and the second are systematic.}
    \label{tab:Kpi}
    \vspace{0.2cm}
    \begin{tabular}{|c|c|c|}
    \hline
    & $\mathcal{B}\ [10^{-6}]$ & $\mathcal{A}_{{\it CP}}$ \\
    \hline
    $B^0\rightarrow K^+\pi^-$ & $20.67\pm0.37\pm0.62$ & $-0.072\pm0.019\pm0.007$ \\
    $B^+\rightarrow K^+\pi^0$ & $14.21\pm0.38\pm0.84$ & $0.013\pm0.027\pm0.005$ \\
    $B^+\rightarrow K^0 \pi^+$ & $24.39\pm0.71\pm0.86$ & $0.046\pm0.029\pm0.007$\\
    $B^0\rightarrow K^0 \pi^0$ & $10.50\pm0.62\pm0.65$ & $-0.01\pm0.12\pm0.05$ \\
    \hline
    $I_{K\pi}$ & \multicolumn{2}{|c|}{$-0.03\pm0.13\pm0.05$} \\
    \hline
    \end{tabular}
\end{table}

\section{Summary}
In summary, we present five new results from Belle II: measurement of $B\rightarrow D^{(*)} K K_S^0$ decays, analyses of $\phi_3/\gamma$ with the GLS and GLW methods, precise measurements of two-body decays that contribute to the determination of $\phi_2/\alpha$, and the $K\pi$ isospin sum rule with a competitive precision to the world's best result.



\end{document}